\documentclass{pasa}

\title[BETA]{The Australian Square Kilometre Array Pathfinder: System Architecture and Specifications of the Boolardy Engineering Test Array}
\author[Hotan et al.]{Hotan, A. W.$^1$\thanks{aidan.hotan@csiro.au}, Bunton, J. D.$^2$, Harvey-Smith, L.$^1$, Humphreys, B.$^1$, Jeffs, B. D.$^3$, Shimwell, T.$^{4}$, Tuthill, J.$^1$, Voronkov, M.$^1$, Allen, G.$^1$, Amy, S.$^1$
% ...
\\
\\
\affil{$^1$CSIRO Astronomy and Space Science, PO Box 76, Epping, NSW 1710, Australia}
\affil{$^2$CSIRO Digital Productivity and Services, PO Box 76, Epping, NSW 1710, Australia}
\affil{$^3$Department of Electrical and Computer Engineering, Brigham Young University, Provo, UT, USA}
\affil{$^4$Leiden Observatory, Leiden University, PO Box 9513, NL-2300 RA Leiden, the Netherlands}
\affil{$^5$SKA Organisation, Jodrell Bank Observatory, Lower Withington, Macclesfield, Cheshire, SK11 9DL, UK}
\affil{$^6$Radio Astronomy Laboratory, University of California Berkeley, USA}
\affil{$^7$Radiation Physics Laboratory, Sydney Medical School, The University of Sydney, NSW, Australia}
\affil{$^8$ Inter-University Centre for Astronomy and Astrophysics, Post Bag 4, Ganeshkhind, Pune University Campus, Pune 411 007, India}
\affil{$^9$International Centre for Radio Astronomy Research (ICRAR)}
\affil{$^{10}$Curtin University, GPO Box U1987, Perth, WA 6845, Australia}
\affil{$^{11}$ARC Centre of Excellence for All-sky Astrophysics (CAASTRO)}
\affil{$^{12}$Sydney Institute for Astronomy (SIfA), School of Physics, The University of Sydney, Sydney, NSW 2006, Australia}
\affil{$^{13}$University of Western Australia, 35 Stirling Highway, Crawley, WA 6009, Australia}}
\jid{PASA}
\doi{10.1017/pas.\the\year.xxx}
\jyear{\the\year}

\usepackage[authoryear]{natbib}
\bibpunct{(}{)}{;}{a}{}{,}
\begin{document}

\begin{abstract}
This paper describes the system architecture of a newly constructed radio telescope - the Boolardy Engineering Test Array, which is a prototype of the Australian Square Kilometre Array Pathfinder telescope. Phased array feed technology is used to form multiple simultaneous beams per antenna, providing astronomers with unprecedented survey speed. The test array described here is a 6-antenna interferometer, fitted with prototype signal processing hardware capable of forming at least 9 dual-polarisation beams simultaneously, allowing several square degrees to be imaged in a single pointed observation. The main purpose of the test array is to develop beamforming and wide-field calibration methods for use with the full telescope, but it will also be capable of limited early science demonstrations.
\end{abstract}
 
\maketitle

\section{INTRODUCTION}
\label{sec:intro}

As our understanding of the radio universe improves, there is an increasing demand for telescopes with faster survey speeds, higher dynamic range, higher time resolution and more sophisticated interference mitigation capabilities. The Square Kilometre Array (SKA\footnote{http://www.skatelescope.org}) \citep{2004NewAR..48..979C,dewdney09} is the key international project tasked with building a radio observatory that delivers these improvements and also surpasses the sensitivity of current radio telescopes by several orders of magnitude. This is only possible if the SKA is constructed in a radio-quiet environment, far from terrestrial sources of interference that would otherwise dominate the received power.

The design of the SKA will be based around a large number of interconnected, mass-produced antennas, with several different technological implementations. In particular, the proposed dish array concept makes use of many 15\,m parabolic reflectors. One advantage of this `large number - small diameter' approach is the potential to image a large simultaneous field of view \citep{2007PASA...24..174J} due to the relatively large primary beam width of small antennas. To further improve field of view, several technological developments are currently being assessed worldwide. These include Phased Array Feeds (PAFs), which replace a traditional wave-guide feed horn with an array of receptors located in the focal plane of a reflector. This allows multiple, independently steerable primary beams to be synthesised electronically. The SKA concept also calls for the use of aperture arrays, consisting of wide-angle receptors mounted on the ground and directed skywards. These offer the potential for even larger fields of view and are particularly suited to low radio frequencies where fewer elements are required to densely sample the collecting area.

In the lead-up to SKA phase one, three precursor telescopes will be constructed at the observatory sites selected for the full SKA. ASKAP\footnote{http://www.atnf.csiro.au/projects/askap/} (of which this paper describes the first stage, see also \citealt{2009IEEEP..97.1507D,sch12}) is the SKA precursor dedicated to verifying PAF technology. It is being constructed by the Australian Commonwealth Scientific and Industrial Research Organisation (CSIRO) Astronomy and Space Science Division, which also operates and manages the Murchison Radio-astronomy Observatory (MRO), which is the Australian SKA observatory site (see Section \ref{sec:mro}). The low-frequency aperture array precursor is the Murchison Widefield Array (MWA\footnote{http://www.mwatelescope.org/}) \citep{2013PASA...30....7T}, which is managed by an international consortium and co-located with ASKAP in Western Australia. The South African precursor, MeerKAT\footnote{http://www.ska.ac.za/meerkat/}, consists of an array of 64 dish antennas with  13.5\,m effective diameter primary reflectors and single pixel feeds \citep{2009IEEEP..97.1522J}.

In addition to the precursors, several other pathfinder projects will act as technology demonstrators at other observatory sites. In particular, LOFAR \citep{2009IEEEP..97.1431D} and APERTIF \citep{2008AIPC.1035..265V} are large-scale projects making use of aperture arrays and PAFs respectively.

ASKAP will be the first radio telescope designed specifically to use PAF technology. Some of the many unknown factors include the best method for forming phased array beams for a particular imaging task, the inherent pattern stability of a phased array beam and the optimal arrangement of multiple beams for combination into a larger field of view. In order to investigate these issues while hardware development and construction of components for the full ASKAP is underway, we have fitted 6 antennas with prototype PAFs and installed a digital system capable of forming and correlating at least 9 dual-polarisation beams from each of these 6 antennas. This prototype array, known as the Boolardy Engineering Test Array (BETA), is now operational. 

The majority of this paper describes the various key subsystems of BETA, beginning with a summary of its overall characteristics and finishing with a discussion of planned commissioning activities and the results of early observations.

\section{THE MURCHISON RADIO-ASTRONOMY OBSERVATORY}
\label{sec:mro}

BETA is located at the MRO, an area of sparsely-populated land in Western Australia. The MRO is 305\,km North-East of Geraldton, the nearest major town situated on the coast of Western Australia. Access to the site is via unsealed road, which also serves the needs of several pastoral lease-holds that operate in the surrounding area. The observatory is protected within a radio quiet zone, with regulation of the radio spectrum within a 260\,km radius of the core, which is located at 26.697$^\circ$S, 116.631$^\circ$E. Several mechanisms are in place to protect the radio quiet nature of the site, including the Western Australian government's mining resources management area, and the Australian government's radiocommunications (Mid-West Radio Quiet Zone) frequency band plan 2011 (F2011L01520).

Major infrastructure works on the site were completed in 2012, including construction of 36$\times$12\,m diameter antennas (6 of which have been fitted with prototype PAFs for BETA), the provision of roads, high-speed data links and a shielded control building. Each of the 12\,m antennas is serviced by underground cables, including a 6.6 kV power feed and 216 single-mode optical fibre cores.  The precise position of one reference antenna (at pad number 29) was determined with respect to the International Terrestrial Reference Frame by long-baseline interferometry \citep{2011PASA...28..107P}. Power for the site is currently provided by two hot-swappable 500\,kVA containerised diesel generators, with a permanent solar-diesel hybrid power station under development.

The control building was designed to shield the observatory from radio emission generated by digital signal processing systems (including the computers and networking hardware inherent in modern interferometers), infrastructure and the indoor activities of site staff. An outer layer provides over 80\,dB of shielding for staff offices, workshops and plant (air conditioning, cooling, water filtration, electrical switch gear, etc.). Enclosed within this larger structure is a second layer of 80\,dB shielding surrounding the digital electronics and computing systems responsible for processing astronomy data.

Cooling of all electronics in the central building is accomplished by the circulation of chilled water, using a hybrid system that can reject the heat generated within the building into a geothermal reservoir or the atmosphere as required for optimum efficiency. Each BETA antenna has an independent water cooling system that serves its PAF and pedestal electronics.

\section{SYSTEM OVERVIEW}

BETA makes use of 6 fully-steerable 12\,m parabolic antennas (See Section \ref{sec:ant}). Each antenna has a PAF (See Section \ref{sec:paf}) at its focus which gives a field of view many times larger than a traditional feed horn \citep{bunton2010achievable}. Radio signals received by each of 188 individual receptors in the PAF are amplified and transmitted (at the sky frequency) to the antenna pedestal over coaxial cable. In the pedestal (See Section \ref{sec:pedestal}), the signals are down-converted to an intermediate frequency band of width 304\,MHz, and passed through programmable attenuators that optimise the power level for input to a set of Analogue to Digital Converters (ADCs), which are also located in the pedestal. The digitised voltages then pass through an oversampled polyphase filterbank \citep{bn04} that produces channels with a width of 1\,MHz. Digitised data from each channel and PAF port are transmitted via optical fibre back to the central building at the MRO, where they pass into digital beamformers  (See Section \ref{sec:central}). Beams are formed independently on each 1\,MHz channel by multiplying the PAF port voltages by a set of complex weights (one for each port) and summing the result. After beam-forming, data pass into a second, fine filterbank that produces 16,416 channels with 18.5\,kHz frequency resolution. These data are then transmitted to a correlator that calculates visibilities for each channel and baseline, integrated to 5\,second time resolution.

Uncalibrated visibility data are transferred via long-distance optical fibre (See Section \ref{sec:network}) to the Pawsey high performance computing centre\footnote{http://www.ivec.org/facilities/pawsey/} in the city of Perth, Western Australia. There, data are passed into the central processor sub-system, which is responsible for calibration, imaging and other science processing. The raw visibilities first enter an ingest pipeline that conditions all data prior to science processing. This initial conditioning step begins with the incorporation of telescope configuration and state information (known as meta-data) from the monitoring and control system. The ingest pipeline is also responsible for flagging data that should not be imaged (for example, cycles that correspond to antennas slewing). In addition to flagging based on telescope meta-data, the ingest pipeline is capable of dynamically inspecting the data and applying flagging based on, for example, a running visibility amplitude threshold (where all channels are flagged when the input exceeds, say, 5$\sigma$ in Stokes V).

After flagging, the ingest pipeline calibrates the data using either a model of the radio sky, or stored information from a previous observation of a calibrator source. For BETA, the data rate is such that the output from the ingest pipeline can be stored to disk as a Common Astronomy Software Applications (CASA) measurement set\footnote{http://casa.nrao.edu/Memos/229.html} for offline analysis during testing and development.

After calibration, data proceed to the imaging pipeline. Deconvolution is performed independently on each beam, after which a linear mosaic is formed to produce an image of the entire field of view. The resulting images are stored in a science archive where they can be accessed by users world-wide. The external interface to this archive is still being designed, but the final product will be compatible with International Virtual Observatory Alliance (IVOA) standards\footnote{http://www.ivoa.net/documents/VOTable/}.

The process of data acquisition is orchestrated by a monitoring and control sub-system known as the Telescope Operating System (TOS) \citep{2010SPIE.7740E..52G}. This sub-system provides fundamental services such as a logging, monitoring data capture and archiving, alarm management, facility configuration data management and the provision of operator displays. The TOS is also responsible for some aspects of hardware safety, by automatically comparing measured diagnostic data to pre-defined nominal ranges and alerting the operator if a problem is detected.

High-level software known as the Executive manages the scheduling of observations via a queue that executes individual scheduling blocks automatically (see Section \ref{sec:computing}).

The fundamental characteristics of BETA are summarised in Table \ref{tab:specs}. Information on the measured sensitivity and expected survey speed can be found in Sections \ref{sec:ant} and \ref{sec:paf}. The remaining sections expand on the information given above and describe the results of early observations.

\begin{table}[h]
\begin{center}
\caption{Key parameters of the BETA telescope.}
\label{tab:specs}
\begin{tabular}{ll}
\hline 
Number of Antennas & 6 \\
Antenna Diameter & 12\,m \\
Total Collecting Area & 678.6\,m$^2$ \\
Maximum Baseline & 916\,m \\
Angular Resolution & 1.3$'$ (see Fig. \ref{fig:psf})\\
Observing Frequency & 0.7 to 1.8\,GHz \\
Simultaneous Bandwidth & 304\,MHz \\
Frequency Channels & 16416 \\
Frequency Resolution & 18.5\,kHz \\
Simultaneous Beams & 9 (dual-pol)\\
Minimum Integration Time & 5\,s \\
\hline
\end{tabular}
\medskip\\
\end{center}
\end{table}

\section{ANTENNAS}
\label{sec:ant}

\subsection{Design and Construction}

The overall design and construction of each antenna was carried out by the 54th Research Institute of China Electronics Technology Group Corporation (CETC54). Each antenna was pre-assembled in a large factory facility before being fully dismantled and packed into three shipping containers for delivery to the MRO site. 

The antennas are 12\,m diameter parabolic reflectors with a quadruped that holds a PAF at the prime focus. They have a focal ratio (f/D) of 0.5 and move on three axes; azimuth, elevation and roll. The antennas can be driven over a continuous azimuth range of -179$^{\circ}$ to 359$^{\circ}$ and an elevation range of 15$^{\circ}$ to 89$^{\circ}$. The roll axis (or parallactic angle axis) rotates the entire reflector about the optical axis with a range of -179$^{\circ}$ to 179$^{\circ}$ \citep{fjk09}. This unique third axis of rotation allows the orientation of the reflecting surface, quadruped and PAF to be kept fixed with respect to the sky and has sometimes been referred to as a ``sky mount''. This design was preferred over an equatorial mount as it was more economical, easier to construct and provided the significant benefit of keeping all beams stationary with respect to the antenna structure when tracking a source across the sky.

The drive system can move at a maximum rate of 3\,deg/s in azimuth and polarisation, and 1\,deg/s in elevation. Due to the finite slew rates, there is a small zone around the zenith through which the telescope cannot track while maintaining a lock on a source. The tracking system has a positional accuracy of 5$''$ and the antennas are rated to operate in wind conditions with gusts up to 60\,km/h. The TOS will automatically stow the array if wind conditions exceed safe operational limits.

The roll axis eliminates the computational load that would have been required to track position angle using beamforming and reduces the impact of any systematic errors resulting from non-circular symmetry in the beam side lobes. This allows for a more uniform field of view and has greatly simplified commissioning tests in the early stages of hardware deployment. It also provides a unique opportunity to characterise beam patterns and polarisation characteristics. For instance, an observation could be repeated with the orientation of the antenna set to a different position angle. This will provide information on systematic trends in instrumental polarisation and hopefully improve the polarisation performance of the full ASKAP array.

For a description of the mathematical formalism of radio astronomical polarisation and instrumental calibration techniques, please see \cite{1996A&AS..117..137H} and \cite{1996A&AS..117..149S}. In addition, several EVLA Memos\footnote{http://www.aoc.nrao.edu/evla/memolist.shtml} (in particular, 131 and 177) have discussed measurements of instrumental polarisation and its impact on the dynamic range of radio interferometers.

\subsection{Measured Performance}

Photogrammetry experiments show that the surface accuracy of the antennas is very good; root-mean-square deviations from the best-fit paraboloid are typically less than 0.5\,mm at all elevations, for all antennas. This will allow efficient operation up to 20\,GHz if necessary in future.

\begin{figure}[h]
\begin{center}
\includegraphics[scale=0.45, angle=0]{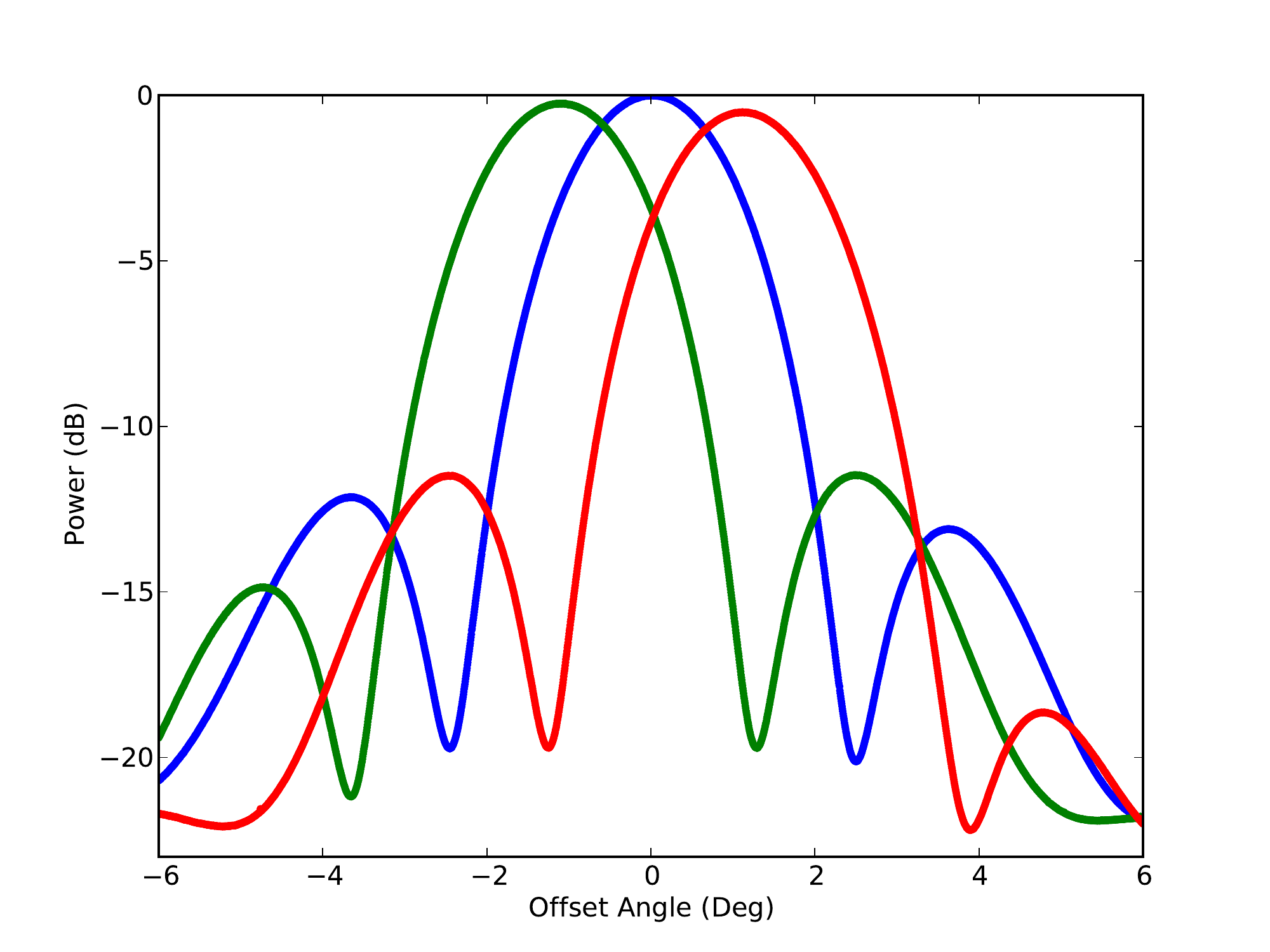}
\caption{Drift-scan cut through a set of three beams formed with maximum sensitivity weights, at a frequency of 860\,MHz over a bandwidth of 1\,MHz}
\label{fig:beam_profile}
\end{center}
\end{figure}

The 12\,m diameter of the primary reflector corresponds to a diffraction-limited beam width at half power of approximately 2\,$^{\circ}$ at the lowest observable frequency, decreasing to 0.8\,$^{\circ}$ at the high-frequency end of the band. Figure \ref{fig:beam_profile} shows a set of beam profiles from one of the BETA antennas, measured using a drift scan through three simultaneous beams formed at 1-degree intervals along the drift axis using the maximum sensitivity algorithm (see Section \ref{sec:paf} for more information). Auto-correlations on the Sun were recorded for a single 1\,MHz channel at a sky frequency of 860\,MHz to estimate the beam profile. This example measurement does not make use of interferometric methods and is intended as a demonstration only. The nature of the formed beams (and indeed, their combined properties over the full field of view) depends heavily on the beamforming technique used, which is still an active area of research. 

It is interesting to note that the off-axis primary beams exhibit an asymmetry in the level of the first side-lobe, which is a side-effect of the maximum sensitivity beamforming algorithm. The imaging software currently makes the assumption that all beams are identical and deviations from this assumption may result in dynamic range limitations. It is possible to constrain the shape of a beam when forming a weight solution, though this will likely be at the cost of sensitivity (see Section \ref{sec:pafCal}). An important use of BETA will be to quantify these trade-offs and establish optimal beamforming methods tailored for specific astronomical survey strategies. Future publications will describe the results of these ongoing studies.

\subsection{Array Layout}

\begin{figure}[h]
\begin{center}
\includegraphics[scale=0.075, angle=0]{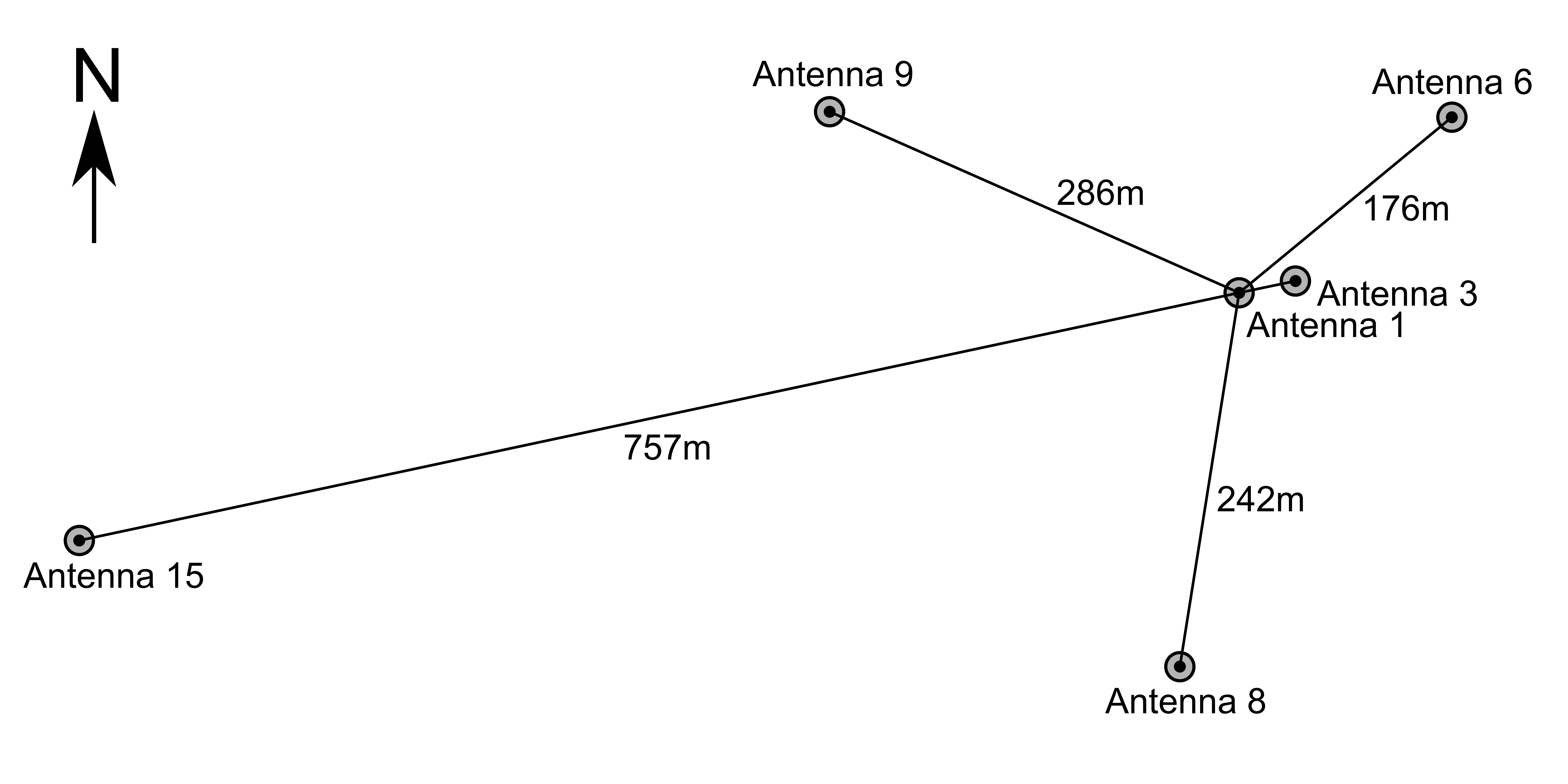}
\caption{Scale diagram of the relative geographic positions of the BETA antennas, with selected baseline distances marked.}
\label{array_map}
\end{center}
\end{figure}

\begin{figure}[h]
\begin{center}
\includegraphics[scale=0.45, angle=0]{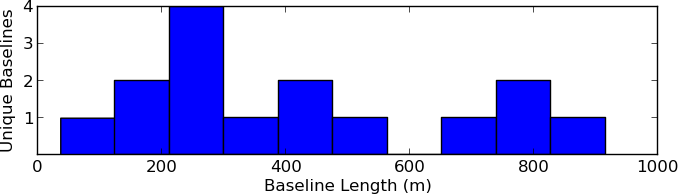}
\caption{Histogram showing the distribution of baseline lengths, with ten bins of 100\,m width.}
\label{fig:baseline_histogram}
\end{center}
\end{figure}

The 6 BETA antennas were chosen to provide a mix of relatively short baselines with small fringe rates for ease of commissioning, and slightly longer baselines to provide spatial resolution. Figure \ref{array_map} shows the relative position of each antenna on a scale diagram. Table \ref{tab:baselines} shows the distance between each pair of antennas in the array and Figure \ref{fig:baseline_histogram} shows a histogram of the baseline distribution. The longest baseline is 916\,m (from antenna 6 to 15) and the shortest is 37\,m (from antenna 1 to 3). 

This layout favours East-West orientations, but the relative elongation of the array in the East-West direction is only a factor of approximately 2, as compared to the North-South direction.

\begin{table}[h]
\begin{center}
\caption{Baseline lengths for all pairs of antennas in metres.}\label{tab:baselines}
\begin{tabular}{c|cccccc}
\hline
ASKAP & & & & & & \\
Antenna &   1 &   3 &   6 &   8 &   9 &  15 \\
Number & & & & & & \\
\hline
  1 &   - &  37 & 176 & 241 & 268 & 756 \\
  3 &  37 &   - & 144 & 257 & 299 & 793 \\
  6 & 176 & 144 &   - & 391 & 380 & 916 \\
  8 & 241 & 257 & 391 &   - & 406 & 706 \\
  9 & 268 & 299 & 380 & 406 &   - & 564 \\
 15 & 756 & 793 & 916 & 706 & 564 &   - \\
\hline
\end{tabular}
\medskip\\
\end{center}
\end{table}

\subsection{Expected Imaging Performance}

\begin{figure}[h]
\begin{center}
\includegraphics[scale=0.35, angle=0]{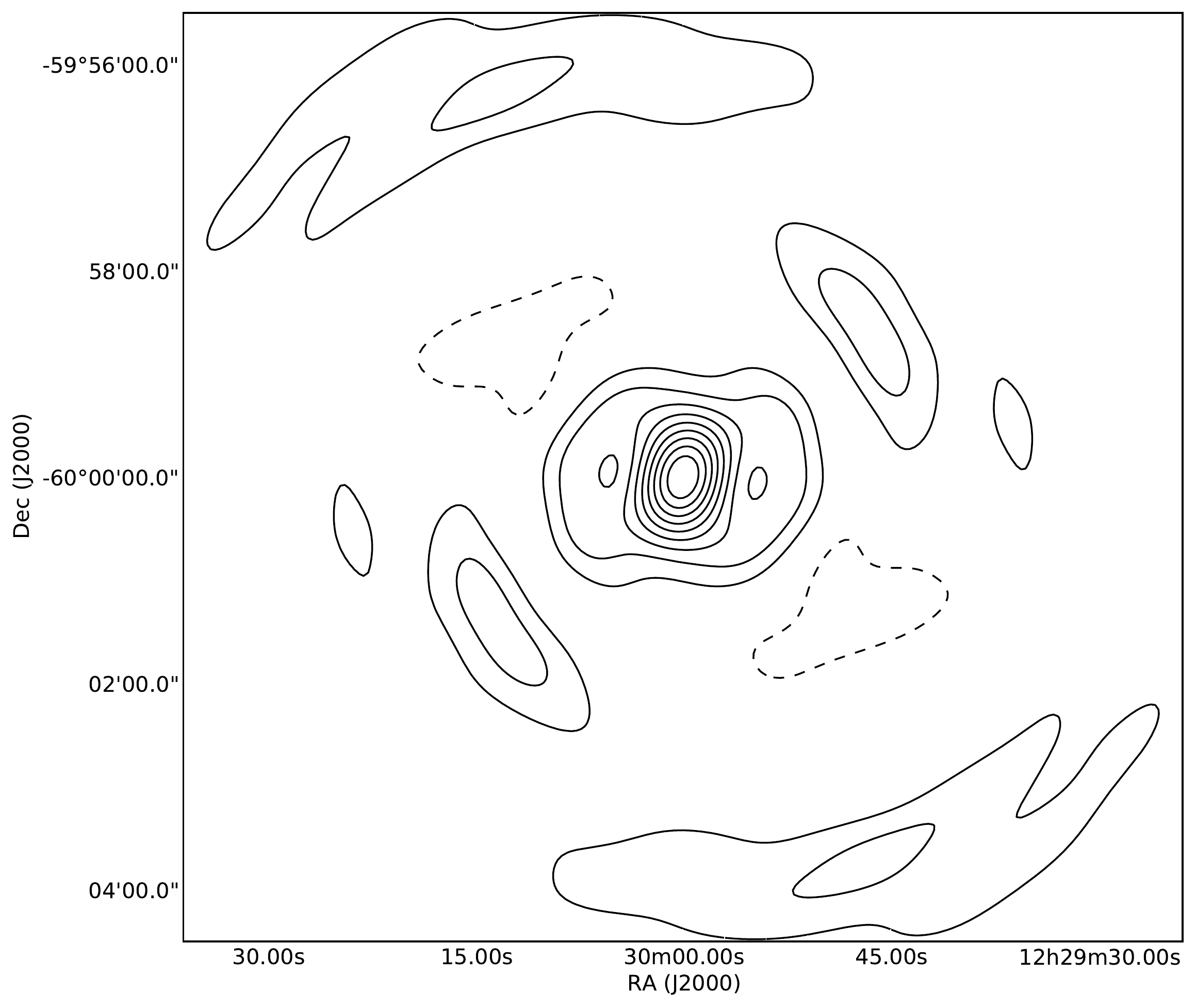}
\caption{Point spread function for a naturally-weighted synthesised beam covering 1.1\,GHz to 1.4\,GHz assuming an 8\,hr observation at a declination of $-$60\,$^{\circ}$. Contours represent a linear scale in 10\% increments, with solid lines being positive and dashed lines negative.}
\label{fig:psf}
\end{center}
\end{figure}

Simulations of a typical 8\,hr observation at a declination of $-$60\,$^{\circ}$ were used to calculate the expected Point Spread Function (PSF) of a single boresight beam (see Figure \ref{fig:psf}). In addition, Table \ref{survey_table} shows predicted survey speeds for BETA, based on the calculations presented in \cite{2007PASA...24..174J}.

\begin{table}[h]
\begin{center}
\small
\caption{Predicted parameters for three different survey modes, split into two columns for the upper and lower parts of the BETA observing band, assuming 9 simultaneous dual-polarisation beams. Sensitivity for the first two rows is expressed as the 1$\sigma$ threshold over the stated bandwidth. The surface brightness survey is characterised by the rate required to reach a 1$\sigma$ limit of 1\,K over the specified bandwidth, assuming the stated spatial resolution which was obtained using a maximum baseline of 400\,m and only 5 of the BETA antennas (antenna 15 excluded due to its relatively long baseline). The final row gives the time required to reach a 1$\sigma$ limit of 1\,mJy/beam for a point source in the field centre.}
\label{survey_table}
\begin{tabular}{lll}
\hline
Frequency, T$_{\rm sys}/\eta$ & 900\,MHz, 75\,K & 1.4\,GHz, 200\,K \\
\hline
Continuum & 1.34\,deg$^2$/hr & 0.29\,deg$^2$/hr\\
 (300\,MHz, & & \\
100\,$\mu$Jy/beam) & & \\
\hline
Line brightness & 1.11\,deg$^2$/hr & 0.24\,deg$^2$/hr \\
 (100\,kHz, & & \\
5\,mJy/beam) & & \\
\hline
Surface brightness & 2.19\,deg$^2$/hr & 0.48\,deg$^2$/hr \\
 (18.5\,kHz, 1\,K) & (170$''$) &  (110$''$) \\
\hline
Point source sens. & 4.03\,min & 18.4\,min \\
 (300\,MHz, & & \\
1\,mJy/beam) & & \\
\hline
\end{tabular}
\medskip\\
\end{center}
\end{table}

\section{PHASED ARRAY FEEDS}
\label{sec:paf}

\subsection{Physical and Electrical Properties}

\begin{figure}[h]
\begin{center}
\includegraphics[scale=0.3, angle=0]{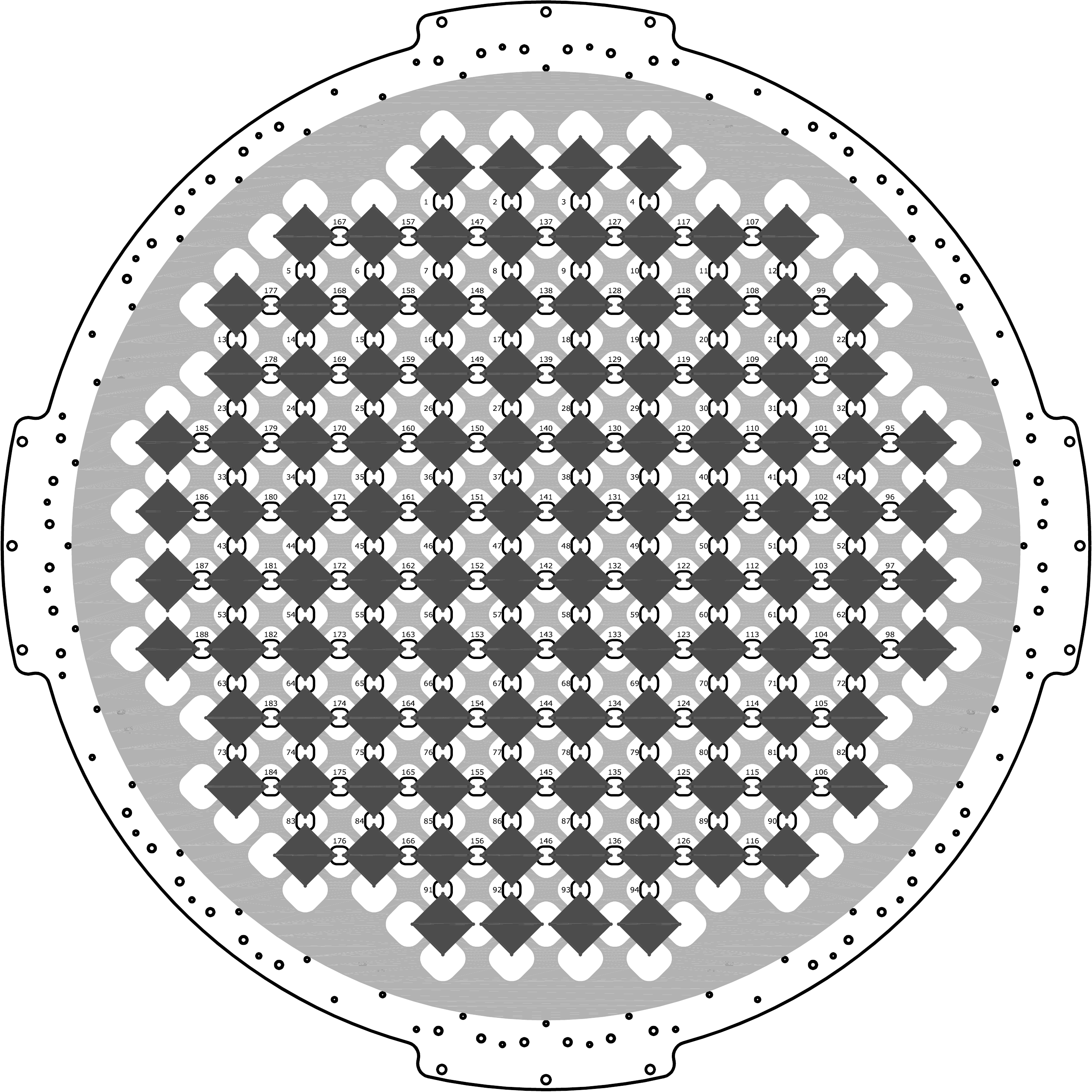}
\caption{Diagram showing the layout of antenna elements on the surface of the ASKAP Mk I PAF. Low noise active baluns are connected at the points of adjacent diamonds (black). The solid aluminium ground plane is shown in outline and the foam spacing between the chequerboard and the ground plane is shaded grey.}
\label{chequerboard}
\end{center}
\end{figure}

Each BETA antenna is equipped with an ASKAP Mk I chequerboard phased array feed \citep{2006ESASP.626E.663H, RDS:RDS5546,IJMOT-2010-11-109, hom11}. These consist of 188 active elements (94 for each of two orthogonal polarisations) arranged in a regular grid (see Figure \ref{chequerboard}) that covers one square metre of the focal plane. Low noise active baluns \citep{shr12} connect between adjacent corners of the chequerboard elements to amplify the signals they receive. As second order intermodulation of strong signals is possible in subsequent amplifiers, selectable sub-octave band filters are included after the first stage. The amplified and filtered signals are then transported by coaxial cable to the antenna pedestal through cable wraps.

Voltages, currents and temperatures in the PAF are monitored, but care is needed to minimise the radio frequency interference that such a monitoring system may produce. This is done by using an optical connection for monitor and control, and an implementation of the Serial Peripheral Interface (SPI) communications protocol, which is only clocked by a master hub when data communications are required.

The feed is housed in a sealed enclosure designed to keep the internal environmental conditions as stable as possible. This enclosure is water-tight and kept at a slight positive pressure by a supply of dry air from the antenna pedestal. Because the unit is sealed, heat generated by the electronics cannot escape directly into the surrounding air. Instead, a water-cooled heat exchange system is used to keep the air within the PAF at a steady temperature of 25\,$^{\circ}$C. Control is achieved by varying the flow rate of cold water through the unit.

The feed was designed to cover a frequency range of 0.7\,GHz to 1.8\,GHz, with a noise temperature of 50\,K, but the prototype chequerboard suffers from an impedance mismatch between the antenna elements and the low-noise amplifiers that results in an elevated noise temperature in the upper part of the band (see Section \ref{sec:beamperf}).

Electromagnetic simulations are critical to the PAF design process and the design of the Mk\,I PAF has allowed improvements in the models used, such that the simulated performance is now a good match to the measured performance. Successive generations of chequerboard currently under development have been verified to meet the 50\,K system temperature specification across the entire band \citep{shr12}. These new Mk II systems will be installed on all ASKAP antennas, eventually replacing the BETA Mk I PAF systems.

In order to fully characterise the performance of the PAF on the sky, it is necessary to form a beam that efficiently illuminates the antenna. Practicalities associated with this are discussed in the next section. 

\subsection{Beamforming and Calibration Methods}
\label{sec:pafCal}

ASKAP belongs to a new class of PAF imaging array instruments where the wide field of view depends on the ability to electronically form many simultaneous, well shaped, and stable beams on the sky.  
Achieving these goals poses (among other issues) two new engineering challenges not encountered in conventional synthesis imaging arrays: digital beamformer weight calculation and beamformer calibration.
The latter is distinct from traditional imaging array calibration.

To keep patterns stable, beamformer weights must be recomputed regularly over intervals ranging from a few days to months in order to adjust for relative per-receptor gain and phase drift, variability in the electronics, and intentional changes in system parameters such as receiver chain gain settings. Much more frequent updates are required if maximum sensitivity is to be achieved for each observation by optimising with respect to the local direction dependent noise field, or for spatial nulling of moving or transient RFI \citep{Landon10,Landon11}.

To-date, all weights used for real data processing on ASKAP and other published PAF results have been from the statistically optimal maximum signal to noise ratio algorithm (max SNR, also known as maximum sensitivity) \citep{applebaum76}
\begin{eqnarray}
y_k [i] & = & {\bf w}_k^T {\bf x}[i] \nonumber \\
{\bf w}_k & = & \hat{\bf R}^{-1}_n \hat{\bf v}_k \nonumber
\end{eqnarray}
where ${\bf x}[i]$ is the vector of complex array voltages for a single frequency channel at time sample $i$, $y_k [i]$ is the $k$th beam time series output, ${\bf w}_k$ is the corresponding vector of complex weights, $\hat{\bf R}_n$ is the integrated sample estimate for the local PAF Array Covariance Matrix (ACM), and $\hat{\bf v}_k$ is the estimated array response vector for a unit magnitude plane wave signal arriving from the direction of the $k$th beam's mainlobe peak.
This approach is optimum with respect to a given noise covariance model $\hat{\bf R}_n$, but beam pattern detail is not directly controlled.
This can lead to pattern variation across beams and over time.

Since the spatial structures of spillover noise, sky noise, and other components can depend on  elevation pointing and other variables, best results are obtained using contemporary $\hat{\bf R}_n$ estimates from a nearby off-source pointing at the same elevation.  
However, it can be challenging to obtain frequent, effectively source-free noise field estimates.
So a single typical $\hat{\bf R}_n$ obtained during a major calibration cycle is often used for a period of several days in computing weights for all beams and dish pointing directions.  The result is a compromise set of  weights which do not truly yield maximum sensitivity for the current observation, but which produce beam patterns that vary little over time.
 
Other algorithms have been proposed which deterministically control the beam pattern.
These can constrain fixed mainlobe or sidelobe shapes (e.g. Gaussian beam or equiripple sidelobes), improve pattern stability, or provide a parametric hybrid trade off between statistically optimum and controlled shape beamforming  \citep{Elmer11,Willis09}.
These advantages come at the cost of reduced sensitivity. 
Also, such methods require a denser, and thus more time consuming beamformer calibration grid observation, which is discussed next.
Evaluation of the relative effectiveness of these candidate beamforming algorithms, particularly as regards to final imaging quality metrics, is a major goal of the ASKAP BETA project.

Beamformer calibration is the process of estimating array response vectors $\hat{\bf v}_k$ for all  beam mainlobe peak directions, {\em and} for all points where deterministic pattern constraints or response samples are desired. For example, constraints could be placed along each beam's -3dB contour to enforce symmetry and size, or in the near sidelobes to set a maximum allowed level. 

Calibration requires a sufficiently compact and bright sky source to excite PAF receptors at high signal to noise ratio so the corresponding array response can be easily extracted from the total signal plus noise array covariance $\hat{\bf R}_{s+n}$.
Array signature $\hat{\bf v}_k$ is estimated as the dominant eigenvector of the generalised eigen equation for $\hat{\bf R}_{s+n}$ and $\hat{\bf R}_{n}$ \citep{Landon10}.
The dish is steered in a grid pattern to place the source at each desired calibration angle relative to dish bore sight.  A few seconds of integration (depending on source brightness) are collected at each  point.
For dense calibration grids needed to map out full beam response patterns, or with many deterministic beam pattern constraints, this process can be time consuming, on the order of hours. 
However, currently for BETA max SNR weight calculations, only nine pointings are needed for the nine real-time beamformer main lobe peaks.  These are recalibrated every few days to account for instrumental drift.  

A significant challenge for ASKAP beamformer calibration is that in the southern sky there are few sufficiently bright sources to serve as reliable calibrators for the relatively small 12\,m dishes.
Two distinct calibration approaches have been used successfully at the Parkes ASKAP test platform to address this problem.
First, we have exploited the large aperture of the co-located Parkes 64\,m telescope to facilitate calibration with weaker sources in the range of a few tens of Janskys.  For this interferometric calibration, the Parkes 64\,m telescope tracks the source while the 12\,m is steered through its grid pattern. The 64\,m telescope's L-band receiver signal is wired into spare ports of the PAF ACM correlator, so cross correlations are computed with all PAF array elements.  Due to the high gain of the 64\,m telescope, the dominant eigenvector of this extended $\hat{\bf R}_{s+n}$ matrix yields the desired  cal vector $\hat{\bf v}_k$.

Second, since the Parkes 64\,m telescope is not always available, and at the MRO there is no large dish, a stand-alone calibration method is needed for 12\,m dishes.   Because of its finite size and intrinsic variability, the Sun is not an obvious candidate for a beam-forming reference, but we have found that it is the only source available in the southern sky with sufficient radio flux for the maximum signal to noise method to work reliably on a single 12\,m antenna and 1\,MHz frequency coarse channel width. 

\begin{figure}[h]
\begin{center}
\includegraphics[scale=0.31, angle=0]{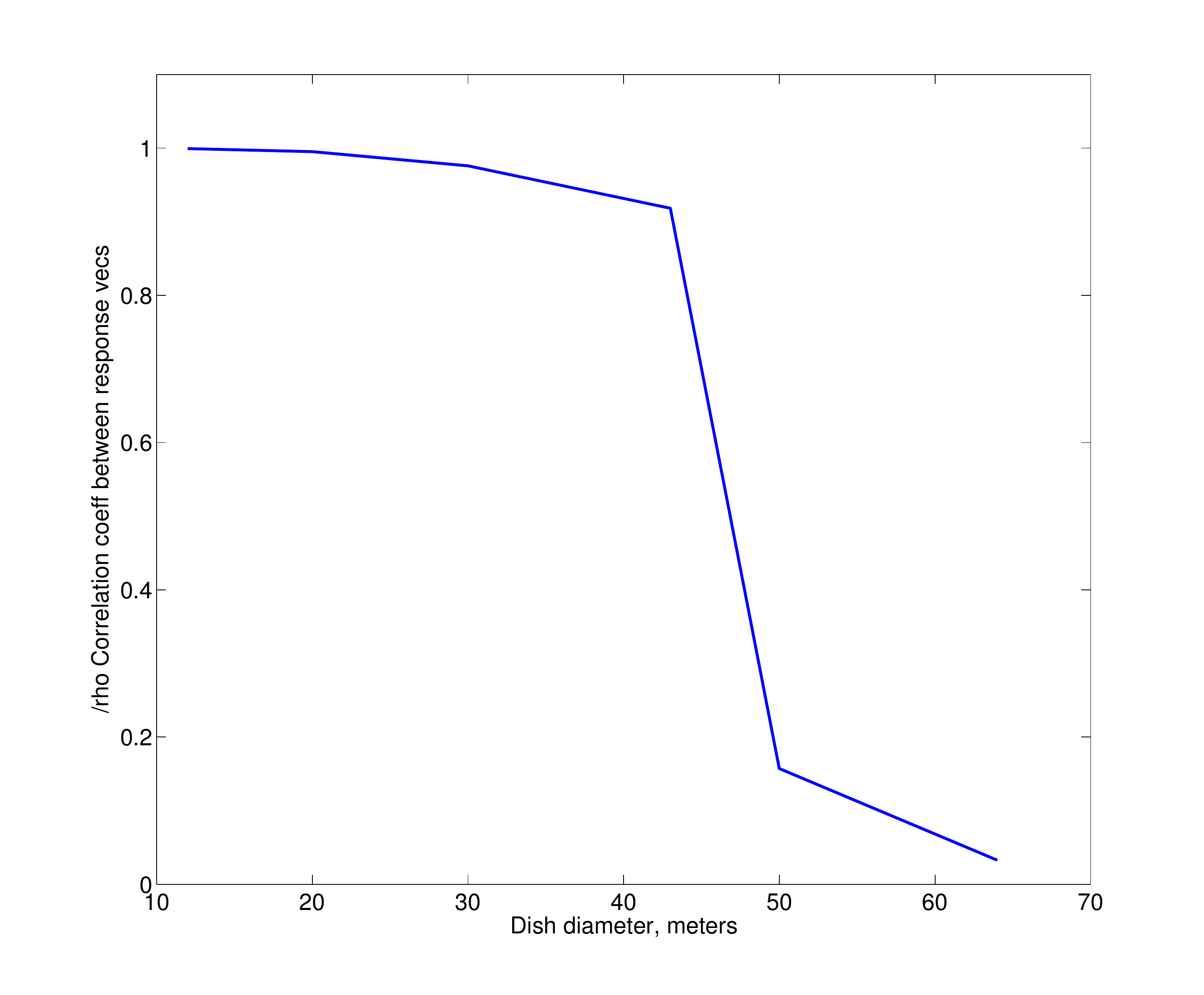}
\caption{Correlation coefficient, $\rho$ between array response $\hat{\bf v}$ estimated with 
the Sun as calibrator, and the true point-source response.}
\label{fig:calCorCoeff}
\end{center}
\end{figure}

To confirm suitability of the Sun as a calibrator, we included a model of its average intensity distribution over the disk and corona regions \citep{Kuzmin66} in a detailed electromagnetic simulation of dish, supports, and array.  
The simulated PAF was the BYU/NRAO 19 element array, which though smaller has similar calibration performance for the central beams as the ASKAP BETA PAF.
Receiver, sky, and spillover noise, along with element mutual coupling effects were carefully represented in the simulation which produced the observed sample array covariance matrices 
$\hat{\bf R}_{s+n}$ and true point-source array response vectors ${\bf v}_k$ as output products.
Estimates $\hat{\bf v}_k$ were then computed as described above from the simulated $\hat{\bf R}_{s+n}$, and the deterministic vector correlation coefficients 
$\rho = (\hat{\bf v}_k^T {\bf v}_k) / ( || \hat{\bf v}_k || \, || {\bf v}_k ||)$ between estimated and true calibration vectors were computed as a function of dish diameter (See Figure \ref{fig:calCorCoeff}).
A $\rho$ value approaching $1.0$ indicates no corruption of the calibration due to the extended non-point-like structure of the Sun.
For the 12\,m dish, $\rho = 0.9994$, and no discernible distortion was found in beam patterns using the Sun calibration vectors for dishes as large as 30\,m.  
This is because the 32.1' diameter Sun is unresolved in the beamwidth of these smaller dishes.
Average intensity variation is not a problem since eigenvectors are arbitrarily normalised, and variation across disc is also unresolved in the natural dish beam.

Optimising the calibration processes and developing methods for extending interferometric calibration to use several co-phased 12\,m dishes as the high gain reference will be major goals of the BETA project.

\subsection{Beamformed System Temperature}
\label{sec:beamperf}

The PAFs used on BETA are the first of their kind and the first ever to be used as part of a synthesis telescope. They are not the final design that will be used for ASKAP, so the measured performance shown here does not reflect the final system configuration. The BETA platform is an engineering prototype and the lessons learnt during its design and construction have already fed back into the design of a Mk II PAF that meets the system temperature requirements for ASKAP across the entire observing band. Here, we describe the as-built measured parameters of the BETA system, knowing that it exhibits a higher than intended system temperature in the upper part of the band.

\begin{figure}[h]
\begin{center}
\includegraphics[scale=0.44, angle=0]{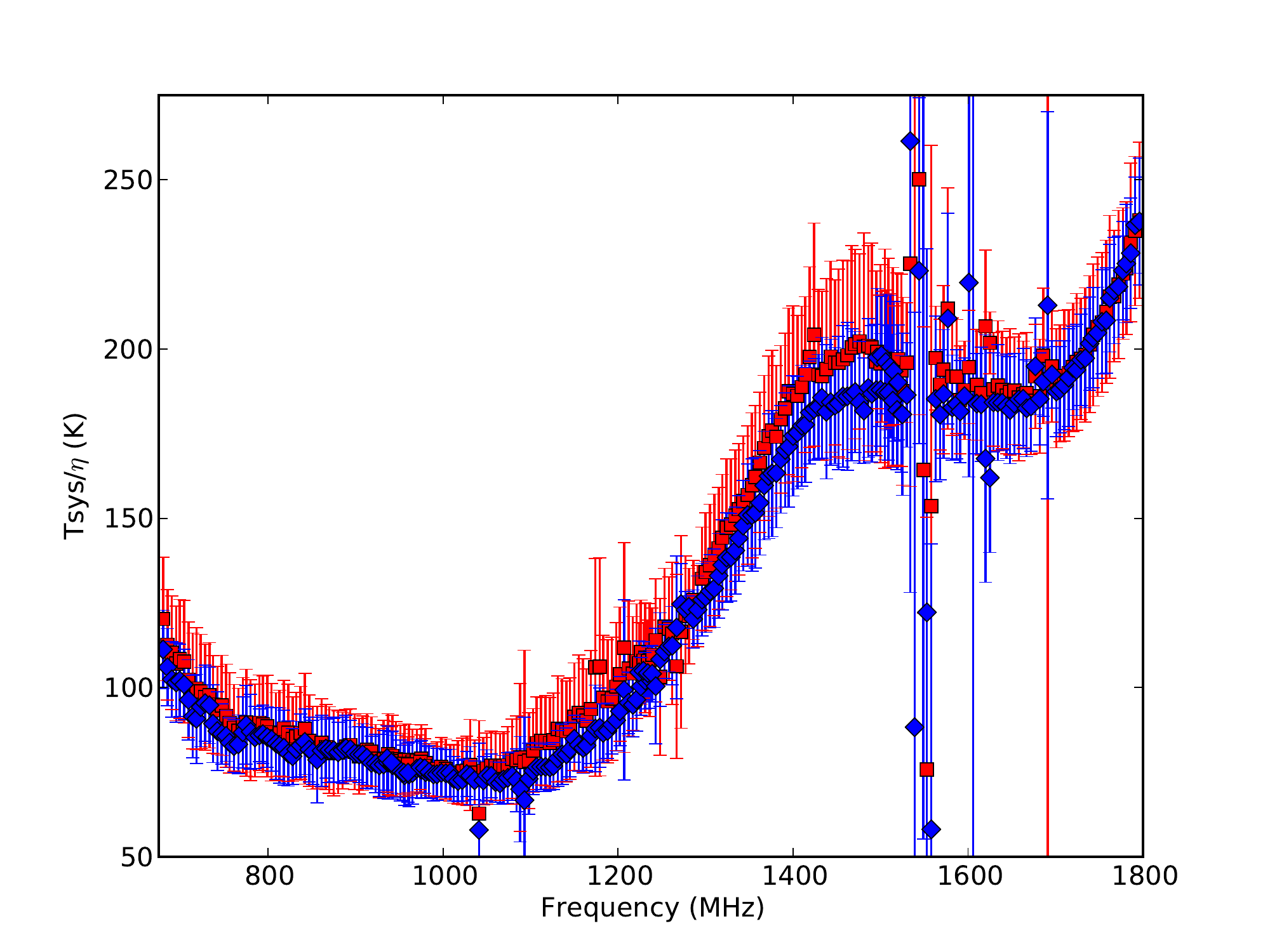}
\caption{Plot of the measured beamformed system temperature of the prototype chequerboard PAF as a function of frequency, with 1\,MHz resolution. Data were recorded using antenna 6 at the MRO. Uncertainty estimates represent 1$\sigma$ limits over four independent trials. Square and diamond markers represent two orthogonal polarisations, recorded simultaneously. A Mk II chequerboard PAF is currently under construction and has already demonstrated significantly better performance.}
\label{fig:tsys}
\end{center}
\end{figure}

Aperture array measurements (where the PAF is operated on the ground, pointing upwards at the cold sky or a warm absorber load held overhead) were used to measure the Y-factor of the PAF, giving its system temperature. On-dish measurements (Figure \ref{fig:tsys}) give the combined T$_{\rm sys}$/$\eta$ (a metric suitable for comparing feed systems independently of collecting area) which by comparison to aperture array measurements suggests that the efficiency factor of a formed beam is approximately 0.7.  A boresight beam was formed on the Sun and then used to observe either Taurus A or Virgo A for flux calibration. 

These measurements use max SNR beamforming on a single dish, performed independently on each 1\,MHz channel, using the Sun as a reference source. The uncertainty in the measurement over several trials suggests that the efficiency of a beam can vary slightly from one beam-forming attempt to the next. Given that the process of Maximum S/N beamforming relies on experimentally-determined information and imposes no constraint on the shape of the beam, this is not surprising. Single-dish beamforming methods are expected to perform poorly compared to methods that use correlations between the antennas, which we are currently developing. The subject of beam optimisation is too large to be covered here and will be the topic of future publications. It will be possible to trade sensitivity for beam symmetry, frequency independence or other desirable properties. Individual science goals will likely require different beamforming optimisations.

Figure \ref{fig:tsys} also shows several channels with unusually large scatter. The worst of these fall between 1540 and 1560\,MHz, corresponding to the service downlink band used by Thuraya 3, a geosynchronous communications satellite that provides mobile telecommunications to the Asia-Pacific region. It is likely that the beam-forming process preferentially steers towards this artificial source when its signal is present in the field of view. We are actively researching methods (such as subspace projection) that allow a spatial null to be placed on known sources of interference when beamforming. This method of interference mitigation is a unique capability of phased array systems.

\section{PEDESTAL ELECTRONICS}
\label{sec:pedestal}

Radio frequency signals from the PAF are down-converted and sampled in the antenna pedestal. These data are processed by a filterbank and then a subset of the frequency channels are sent to the central site building over optical fibre. In order to minimise the interference generated by the pedestal electronics, these systems are housed in shielded cabinets that provide isolation between the analogue and digital domains, and also isolation from the outside world.

\begin{figure}[h]
\begin{center}
\includegraphics[scale=0.48, angle=0]{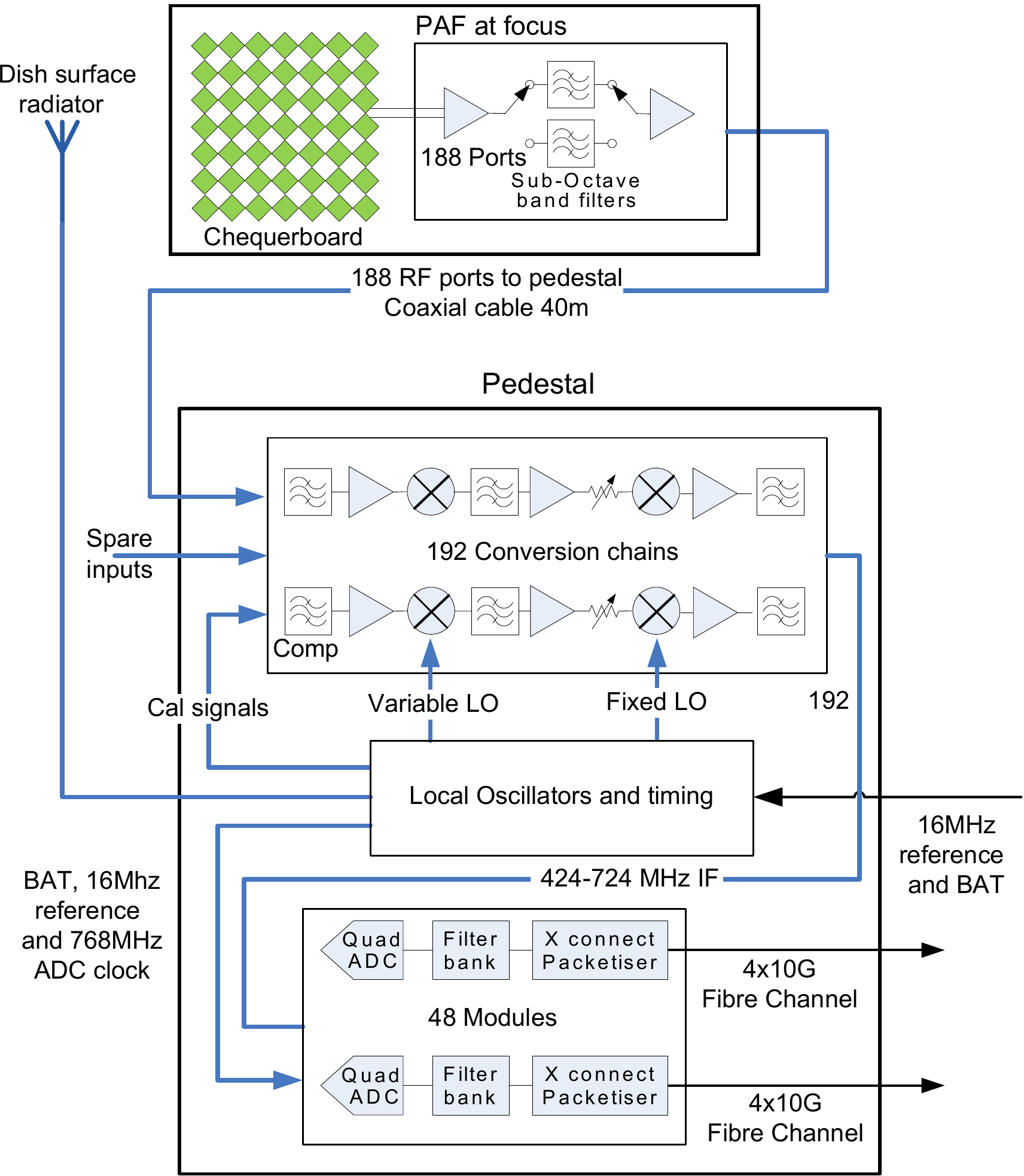}
\caption{Diagram of hardware components and signal flow in the pedestal at each antenna.}
\label{fig:pedestal}
\end{center}
\end{figure}

Figure \ref{fig:pedestal} shows the major components of the signal path contained within the antenna structure.

\subsection{Analogue Stage}

Within the PAF, the signal from each active balun passes through a band-pass filter that selects frequencies in the range 700-1800\,MHz. The band-limited signal is further amplified and then passed through a switch-able band-select filter. Two overlapping ranges are available, 700-1300\,MHz and 1000-1800\,MHz. The output from each filter is amplified again and transmitted at the sky frequency via coaxial cable, through a series of mechanical twister mechanisms that ensure safe passage through the rotating structure of the antenna, to the input side of the shielded cabinet in the pedestal.

In the analogue cabinet all the Radio Frequency (RF) signals undergo a dual-heterodyne down conversion to an Intermediate Frequency (IF) band of 422 to 726\,MHz. This is suitable for the second Nyquist zone of the analogue to digital converter, which is clocked at 768\,MS/s. IF power levels are monitored, as well as voltage, current and temperature for key components. The IF signals pass through variable attenuators with 1\,dB steps and 15\,dB range, which are used to ensure an appropriate input level for the analogue to digital converters.

The local oscillators used for down conversion are derived from a 16\,MHz reference signal supplied from the central building. Presently, the reference signal is derived from a Rubidium clock, which can be locked to a neighbouring hydrogen maser for additional phase stability. On the receiving end, integer-N Phase Locked Loops (PLLs) with purely integer frequency multiplication factors are used in each pedestal to keep the local oscillators in phase with the reference signal. This ensures that phase noise is uncorrelated between each individual antenna. As the PLL is limited to integer multiples of 16 MHz, setting of the band centre frequency is in 16 MHz steps. Limiting the PLL to integer multiples of 16\,MHz also ensures all antennas have the same phase relation after a restart.

\subsection{Digital Data Processing}

The signals pass from the analogue cabinet into the digital cabinet, which holds four racks of CSIRO-designed DragonFly-2 circuit boards \citep{beamformerPaper}. Each DragonFly-2 board has four ADCs, a Field Programmable Gate Array (FPGA) for signal processing and interfacing and four 10\,Gb/s single mode optical links to transport data to the central building. The sampler clock is obtained from the LO subsystem and in addition, the 16\,MHz reference and a precise measure of Coordinated Universal Time (UTC) encoded as a digital signal (referred to as Binary Atomic Time, BAT) are supplied to each DragonFly-2. This allows a unique time to be assigned to any sample from the ADC. The signals from the digitisers then pass through an oversampled polyphase filterbank \citep{bn04} with a frequency channel spacing of 1\,MHz and an oversampling ratio of 32/27. The desired bandwidth for each channel is 1\,MHz and oversampling is used to reduce aliasing \citep{schafer73,crochiere80,wilbur04,filterbankPaper}. The 1\,MHz channel width was chosen to ensure that a narrow-band (weight-and-sum) beamformer \citep{665} can be used without smearing the beams.

The digitiser continuously calculates sampling statistics (the probability distribution function of the output) and makes this information available to the central control system. In the output of the polyphase filterbank, the power levels in each channel are also monitored. Attenuators in the analogue conversion system provide a 15\,dB programmable range that is automatically set to allow for on-line scaling of the ADC input. This is necessary to balance the conflicting requirements of maintaining adequate signal to noise ratio in the quantised astronomical signals while reducing the possibility of ADC saturation due to strong interference.

Four inputs to the down conversion subsystem (in addition to the 188 inputs from the PAF) are reserved for PAF calibration signals and future developments; one possible example being a reference antenna for active interference mitigation. The PAF calibration signal consists of low-power broad-band noise radiated from the surface of the dish. A copy of the calibration signal is sent directly to the receiver and correlated with the PAF outputs, allowing real-time monitoring of the relative gain and phase for each element. This information can in turn be used to adjust beam weights for maximum pattern stability \citep{hcr+10}. The power of this noise source is kept to an absolute minimum to reduce the possibility of it correlating between adjacent antennas. It is also possible to remotely control the noise source so that brief measurements may be taken over longer intervals of time, further reducing the impact on the astronomical signal.

\section{CENTRAL ELECTRONICS}
\label{sec:central}

\begin{figure}[h]
\begin{center}
\includegraphics[scale=0.5, angle=0]{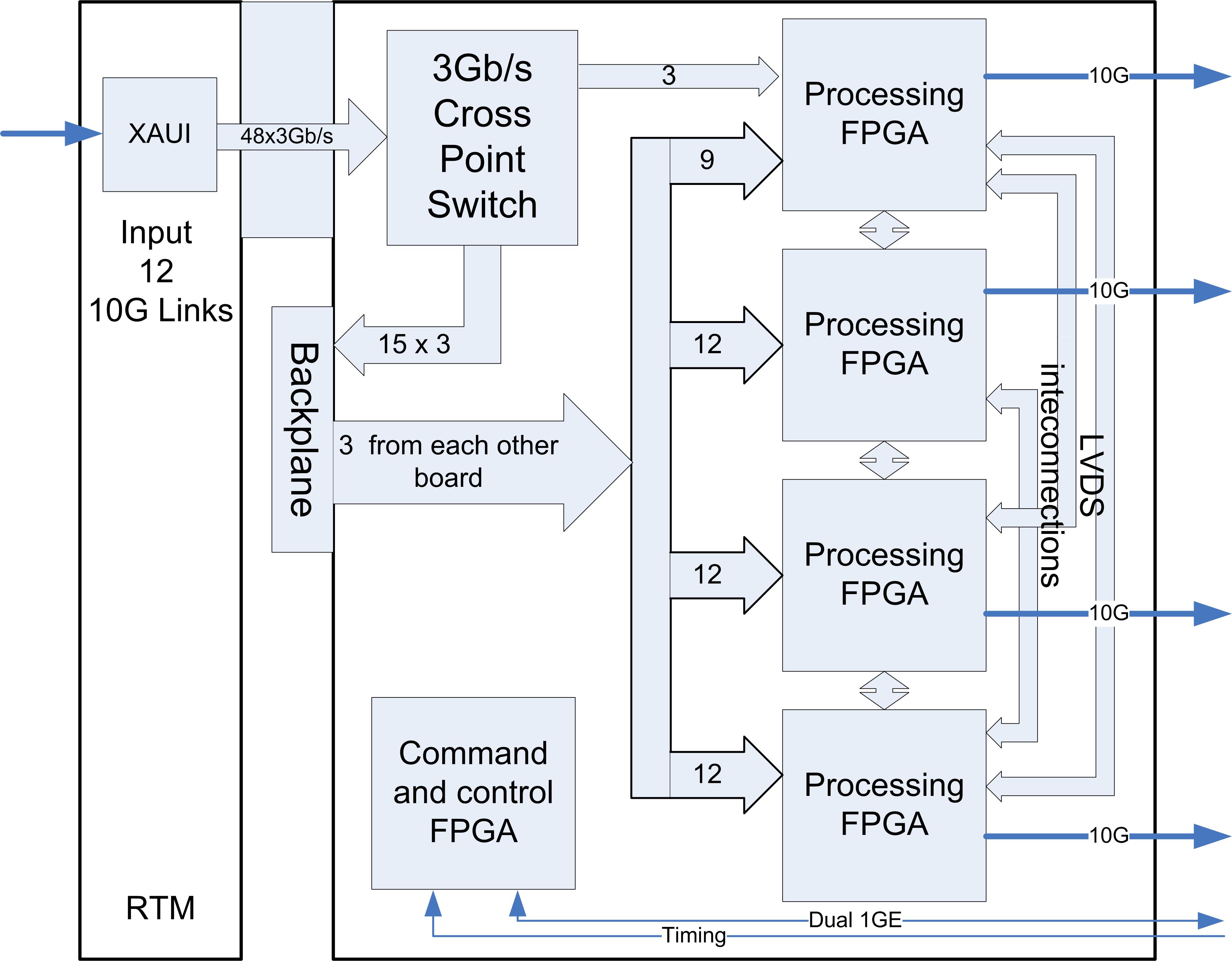}
\caption{Schematic diagram showing the key components of a single Redback-2 system. Signal processing is performed by FPGAs, which are fed data via a 10\,Gb Attachment Unit Interface (AUI) that resides in the RTM tasked with converting signals from optical to electrical form. Communications between FPGAs are handled via Low-Voltage Differential Signalling (LVDS).}
\label{fig:redback}
\end{center}
\end{figure}

The digital data from 304 individual 1\,MHz channels for each PAF port and all 6 antennas are processed in the central building by a common, CSIRO-designed, Redback-2 circuit board \citep{beamformerPaper,dgsXcell}. The basis for the system is the Advanced Telecommunications Computing Architecture (ATCA) standard which specifies a shelf and backplane configuration. The Digital Signal Processing (DSP) hardware for each antenna consists of 16 Redback-2 boards which are inserted into the front of the ATCA shelf and connect through the ATCA backplane and power bus. Rear-mounted boards called Rear Transition Modules (RTMs) provide optical-to-electrical conversion of the incoming data, which are transported over 10\,Gb/s optical fibre links. Each RTM is paired with a DSP card which routes the incoming data onto the ATCA backplane full-mesh cross-connect via a cross-point switch. Each Redback-2 board contains 4 Xilinx Virtex 6 FPGAs for signal processing and a control FPGA to manage communications (see Figure \ref{fig:redback}). Command and control of the beamformer is via an Ethernet connection to the control FPGA on each Redback-2 board. Requests are sent as Universal Datagram Protocol (UDP) packets; the control FPGA decodes a request, executes it and replies with a UDP packet of its own. This packet can be a simple acknowledgement or it may return requested data. The control FPGA does no high level decision making; this is handled by the TOS (described in Section \ref{sec:tos}).

Both the beamformer and the correlator use Redback-2 hardware, although the cards installed in the correlator chassis do not need their high-speed optical output modules as the data rate is sufficiently low that it can be sent via 1\,Gb/s Ethernet over twisted pair copper cables. The use of a common DSP platform also permits some flexibility in managing spare processing cards to keep the overall BETA system operational. However, since only prototype quantities of the DSP cards were produced for BETA, it may be necessary to reduce system capability in the future if DSP card shortages are experienced.

\subsection{Digital Beamformer}
\label{sec:digBeamfmr}

The beamformer system operates on each 1\,MHz channel, producing 9 dual-polarised beams on the sky for each frequency channel. Future expansion to 36 dual-polarised beams is planned for ASKAP. Since there are 304 separate 1\,MHz channels, there are also 304 instances of the beamformer computational unit which are physically distributed across 64 FPGAs on the Redback-2 boards. In the 9-beam system, all 188 PAF ports are used to form the beam. This reduces to an average of 63 in the 36-beam implementation. The data on the links from the digital receivers at the antenna arrive ordered with one quarter of the frequency channels for four ports, whereas each beamformer unit requires all ports for a single frequency channel. The large cross-connect needed to supply data to the beamformer units is realised by the combined routing capabilities of a cross-point switch on each Redback-2, the ATCA backplane full-mesh, interconnections between the Redback-2 FPGAs and within the FPGA itself.

The beamformer operation reduces data throughput and the processing load of the correlator is also significantly decreased because only corresponding beams must be correlated, instead of all PAF elements. The capacity of the beamformer is limited by the number of clock cycles between new data values and the number of beamformer multipliers. Within these constraints the beamformer needs to flexibly allocate the number of complex-multiply-accumulate (CMAC) operations to the beams. This number can vary from beam to beam. 

The beam-forming operation is followed by a fine filterbank that further channelises the signal to the final required frequency resolution. The fine filterbank is a critically-sampled polyphase filter which channelises each 1\,MHz band into 54 fine frequency channels that are 18.52\,kHz wide. Oversampling in the coarse filterbank results in data outside the 1\,MHz band, which are discarded. The fine filterbank is operated in block mode to minimise FPGA memory usage. Each 1\,MHz beamformer unit produces 9 sets of beam pairs at a time for one sample of the 1\,MHz coarse filtered data from all PAF ports. The fine filterbank, however, requires blocks of consecutive samples for one single dual-polarisation beam. A data ``corner-turn'' is therefore necessary between the two modules so that contiguous blocks of samples for a given dual-polarisation beam are presented to the filterbank inputs. Since the corner-turn requires significant storage resources, external DDR3 SDRAM (Synchronous Dynamic Random Access Memory) is used. Associated with the fine filterbank is fine delay tracking (integral-sample delays are removed using a buffer in the DragonFly-2), and fringe stopping.

\subsection{PAF ACM Correlator}

\begin{figure}[h]
\begin{center}
\includegraphics[scale=0.37, angle=0]{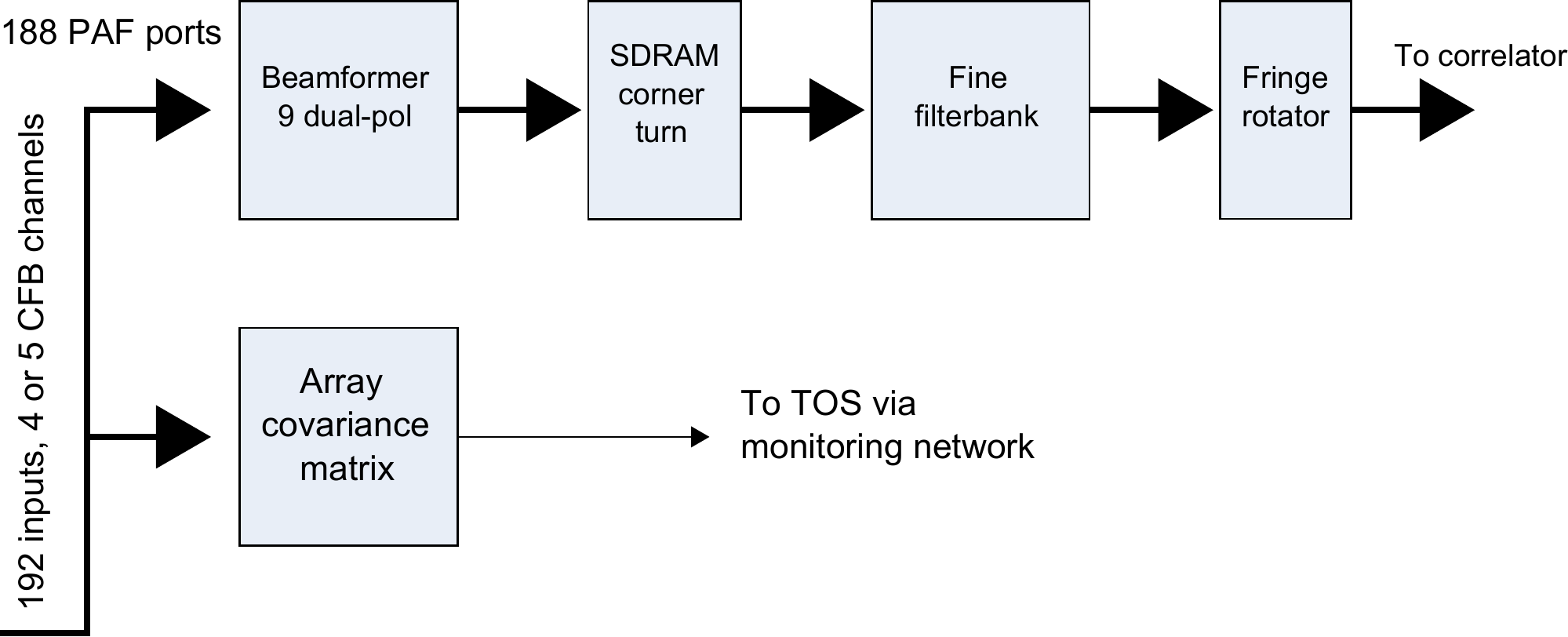}
\caption{Key signal processing tasks performed by the BETA beamformer FPGAs.}
\label{fig:beamformer}
\end{center}
\end{figure}

The beamformer also generates data needed to calibrate the phased array feed. This consists of correlations between calibration signals and PAF port signals and the ACM for the port signals. The compute load of the full ACM is much higher than the beam-forming itself and approximately equal to that of the full correlator that follows the beamformers. For this reason, the ACM module operates with a 25\% duty cycle on a subset of the 1\,MHz channels. All calibration data are sent to the TOS via User Datagram Protocol (UDP) packets. The TOS then runs the algorithms to determine the weighting coefficients needed for each beam, and uploads these to the hardware as necessary. A summary of the key signal processing steps performed in the beamformer FPGAs is shown in Figure \ref{fig:beamformer}.

It is also possible to access critically-sampled data from several beams at the output of the beamformers via UDP over 1\,Gb Ethernet, but only at about 10\% of the full duty cycle and 1/4 the bandwidth. This mode of operation allows for software-based correlation as an alternative to the main data path, which has been useful for initial imaging, system debugging, delay searching and low-level verification of FPGA signal processing firmware.

\subsection{Central Correlator}

The links between the beamformer and correlator are also 10\,Gb/s optical connections.  As the payload is different to that from the digitiser, this stage uses a different but related communications protocol to the pedestal-central building link. The beamformer data that arrives at the Redback-2 correlator boards is redistributed amongst the processing FPGAs so that each FPGA processes a number of fine frequency channels with data from a single dual polarisation beam at one time.

\begin{figure}[h]
\begin{center}
\includegraphics[scale=0.24, angle=0]{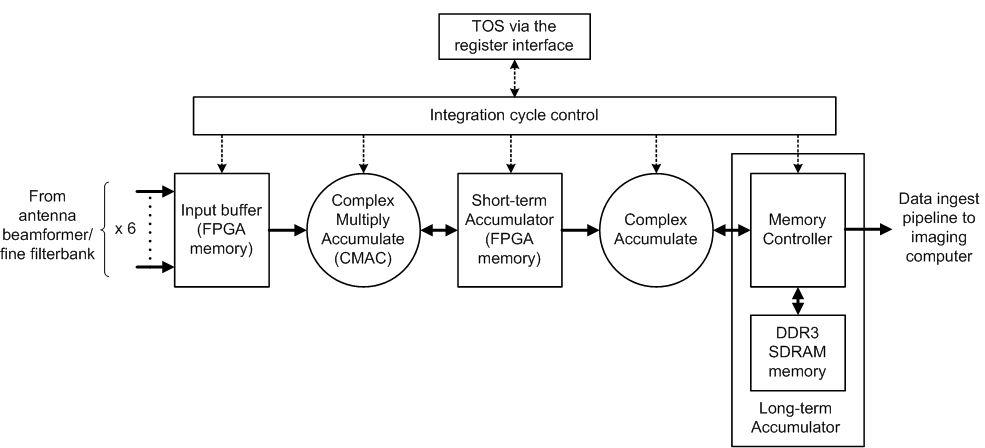}
\caption{Diagram of operations performed in a single correlator cell.}
\label{fig:corrcell}
\end{center}
\end{figure}

At any one time the fine filterbank in the beamformer processes up to 64 coarse 1\,MHz channels for a single dual polarisation beam and each coarse channel generates 54 fine channels. The correlator processes all these data as they are generated.  As the input to the correlator switches between beams and coarse frequency channels, intermediate correlation values are accumulated in external SDRAM. Data in the SDRAM are triple buffered with the buffers switching every 5 seconds. A summary of the signal path is shown in figure \ref{fig:corrcell}.

After accumulation, data are dumped to the external correlator control computer via the 1\,Gb Ethernet connection available on the Redback-2 boards. A network switch is used to aggregate all data onto 10\,Gb Ethernet before being ingested by the control computer. This computer formats the data into a form suitable for transmission to the remote supercomputing facilities in Perth.

\subsection{Data Rate}

The total data output of the BETA correlator is summarised as follows. For the cross-correlations we have 6 antennas producing $6 \times 5/2 = 15$ individual baselines. Each baseline produces 9 dual-polarisation beams and each beam produces $304 \times 54 = 16,416$ fine frequency channels with 4 stokes parameters per channel. This is a total of $15 \times 9 \times 16,416 \times 4 = 8,864,640$ complex numbers.

For the auto-correlations we have 6 individual antennas with 9 dual-polarisation beams per antenna. Each of these produces 16,416 fine frequency channels with 2 auto-correlations for the orthogonal polarisations and 1 cross-polarisation product. This is a total of $6 \times 9 \times 16,416 \times (2+1) =  2,659,392$ complex numbers. 

In addition to the products above, a 64-bit integration count is recorded for each beam and fine frequency channel every cycle, so the data can be normalised by the number of samples integrated. This adds another 147,744 64-bit numbers every 5 seconds.

With an integration cycle of 5 seconds and a long-term accumulator word-length of 32 bits, the total data output rate from the correlator is $(8,864,640 + 2,659,392 + 147,744) \times (32 \times 2)$ bits every 5 seconds, or 142.5\,Mb/s (using 1024$^2$ to convert to Mb).

The ingest pipeline combines these values with appropriate meta-data to define a CASA measurement set. The addition of meta-data causes a small increase to the overall data rate, resulting in a final figure of 153.9 Mb/s reaching the disk. This means that a 1-hour observation with 5\,second integrations and 18.5\,kHz frequency resolution requires 68\,GB of storage space. Consequently, a 12-hour observation produces 816\,GB or 0.8\,TB of data.

\section{NETWORK INFRASTRUCTURE}
\label{sec:network}

There are several different networks within the ASKAP system, providing connectivity from the antenna pedestals to the central building, then a local area network within the central building, and a long-haul network link between the MRO and the Pawsey centre in Perth.

Between each antenna and the central building are 216 single mode fibres. Of these, 188 (one for each PAF element) carry digitised RF data via 10\,Gb/s unrouted connections between the DragonFly-2 digitiser cards and the Redback-2 beamformer cards. A small number of the remaining fibre cores are used for voice and data communication on a routed monitoring and control network.

Within the central site building there exists two separate local area networks, a general purpose network and a science network. The general purpose network hosts IP telephones, laptops, workstations, monitor and control (including links to the antenna pedestals) and environmental monitoring devices. The science network carries the output from the correlator to the correlator control computers, and then finally to the multiplexer terminal that provides connectivity to Perth.

The long-haul network to Perth consists of two major networks, one from the MRO to Geraldton, and the second from Geraldton to Perth. The network from the MRO to Geraldton was constructed by CSIRO and AARNet\footnote{http://www.aarnet.edu.au} as part of the ASKAP project. This network requires three repeater stations, located in Mullewa and at the Yuin and Murgoo pastoral stations.

The Geraldton to Perth link uses a public network constructed by the federal government as part of the regional backbone blackspot program. CSIRO has an access agreement granting several wavelengths to the ASKAP project, allowing a dedicated layer-2 network between the MRO and the Pawsey centre.

\section{COMPUTING}
\label{sec:computing}

\subsection{Monitoring and Control}
\label{sec:tos}

The architecture of the TOS is discussed in detail elsewhere \citep{2010SPIE.7740E..52G} so only a brief outline will be given here. The software is built on a set of tools and libraries collectively known as EPICS\footnote{http://www.aps.anl.gov/epics}. It is designed specifically to support remote operations and large area surveys through a model of automated service observing.

A high-level system known as the Executive is responsible for the execution of scheduling blocks, effectively orchestrating the various hardware and software systems during data acquisition. The implementation of scheduling blocks provides a mechanism for declaring complex observations via the combination of scheduling block parameters and a script, known as the observation procedure. The observation procedure is implemented using Python and exposes telescope control and monitoring interfaces via an application programming interface (API) known as the observation procedure library (OPL).

Expert users, or those performing novel observations, can exploit the flexibility provided via the procedure scripting capability. However, common observing modes are provided via scheduling block templates, providing a pre-tested procedure (script) and requiring the user only specify basic parameters. 

\subsection{Image Processing and Calibration}
\label{subsec:imaging}

The computational challenges associated with interferometry over a wide field of view are known to be significant \citep{2012SPIE.8500E..0LC}. Since BETA utilises only one sixth the total number of ASKAP antennas (and therefore has many fewer and shorter baselines), and one quarter of the number of beams, it will provide a tractable benchmark on the path to a full ASKAP imaging pipeline, at the expense of sensitivity and a more complicated PSF (see Figure \ref{fig:psf}). 

Calibration, imaging, and other science processing is carried out on the ASKAP Central Processor. The hardware platform is a Cray XC30 supercomputer consisting of 472 compute nodes. Each node contains 64 gigabytes of main memory and two Intel Xeon E5-2690 v2 CPUs, each with 10 processing cores. All compute nodes have access to a global Lustre filesystem with a capacity of 1.4 petabytes and capable of sustained I/O rates of 25 gigabytes per second. The Lustre file system provides a buffer for datasets not yet processed as well as for images and other data products that are yet to be archived.

As existing data reduction packages were neither capable of dealing with the high data-rates nor the wide-field imaging challenges of a PAF-equipped interferometer, a custom data reduction package has been developed. Visibilities are gridded with AW-projection and weighted using a single-pass Wiener filter. Continuum images are deconvolved using the multi-scale, multi-frequency-synthesis algorithm described in \cite{rc11}. For spectral line cubes, multi-scale CLEAN \citep{cor08} is used instead.

For BETA a more traditional calibration approach is used, where individual beams are independently calibrated by observing a single reference source. While time-consuming, this method is well tested on other telescopes and allows for bootstrapping while the sensitivity of the array is low due to the small number of available antennas.

\subsection{Data Archiving}

Long term archival storage is provided by a pair of tape libraries (supplied by Spectra Logic) in the Pawsey centre. Two copies of all data products are stored, one copy in each library, to minimise the possibility of data loss. Initially 10 petabytes of tape storage (physically 20 petabytes due to mirroring) is allocated to the ASKAP telescope, with future expansion possible as the libraries are each able to support up to 50 petabytes with the addition of tape media.

This storage pool is used by two different ASKAP archives. An internal archive, known as the {\it Commissioning Archive}, provides storage for any file-based data product the engineering and commissioning teams may wish to store. A public facing archive, the CSIRO ASKAP Science Data Archive (CASDA), is currently being developed. This archive will provide access to data products via a web interface as well as via virtual observatory protocols as defined by the IVOA.

The data rate at the output of the BETA correlator is sufficiently small (compared to the full ASKAP) that it is possible to permanently archive all visibilities. This information is stored in the form of CASA measurement sets. There is sufficient storage space at the MRO to buffer approximately one week of typical data, in order to cope with any network outages.

\subsection{Science Processing}

The focus during the operational life of BETA will be on testing beam-forming and calibration techniques, the imaging pipeline and source-finding algorithms \citep{2012PASA...29..371W} with real-sky data. Due to the large dataset sizes, the full ASKAP science processing system has the difficult task of performing calibration and imaging in a  semi-autonomous mode. This challenge can only be met by gaining an in depth understanding of the telescope and its behaviour under a variety of conditions. Experience gained with BETA will be critical input into this process.

Users of the BETA telescope have access to the Pawsey central processor facility, which can be used to perform data reduction in a traditional iterative and somewhat interactive fashion. As experience is built, automated processing pipelines will be deployed as a demonstration of the ASKAP quasi-realtime processing model. In addition to data reduction with the ASKAP-specific toolset, subsets of data are relatively portable and existing CASA tools are often sufficient for processing tasks. Full continuum datasets (with 1\,MHz channel resolution) are transportable outside of the Pawsey centre, as are small subsets of the spectral line datasets. This allows cross-verification of processing outcomes with a variety of existing systems and software packages.

\section{COMMISSIONING}

The commissioning process involves a comprehensive set of trials ranging from acceptance tests on each individual antenna to imaging tests with all 6 antennas combined. Low-level tests are largely complete; these included verifying the water cooling systems and steady-state environmental conditions within the cabinets while under load, measuring pointing corrections for the antennas and ensuring the correct routing and integrity of the signal from each PAF element.

First fringes between antennas were found using a software-based correlator that operated on small, 1-second bursts of data recorded directly from the output of the beamformer at full time resolution. This allowed a detailed study of signal synchronisation across the array and simplified the integration of the hardware-based correlator. More recently, phase closure was obtained on two different sub-arrays, consisting of antennas 1, 3 and 6, and also antennas 8, 9 and 15, using the hardware-based correlator system. After fringes were obtained on an astronomical source, long tracks were used to verify the initial baseline model and the fringe rotator subsystem. After showing that array phases were relatively stable over periods of 12 hours, test images were made (see below). The final stage of system integration was the configuration of the correlator for the full 6-antenna, 15 baseline mode. This was left until last so that the two independent sub-arrays could be commissioned and debugged in parallel to speed the overall process.

The final test of the full system will be to image two reference fields (covering the Fornax cluster and Circinus galaxy) that have been observed for comparison purposes at the Australia Telescope Compact Array.

\subsection{First Observations}

Preliminary observations using a single pixel feed to characterise the first assembled antenna have already led to interesting scientific outcomes, including the very long baseline imaging project described in \cite{2010AJ....140.1506T}.

\begin{figure}[h]
\begin{center}
\includegraphics[scale=0.48, angle=0]{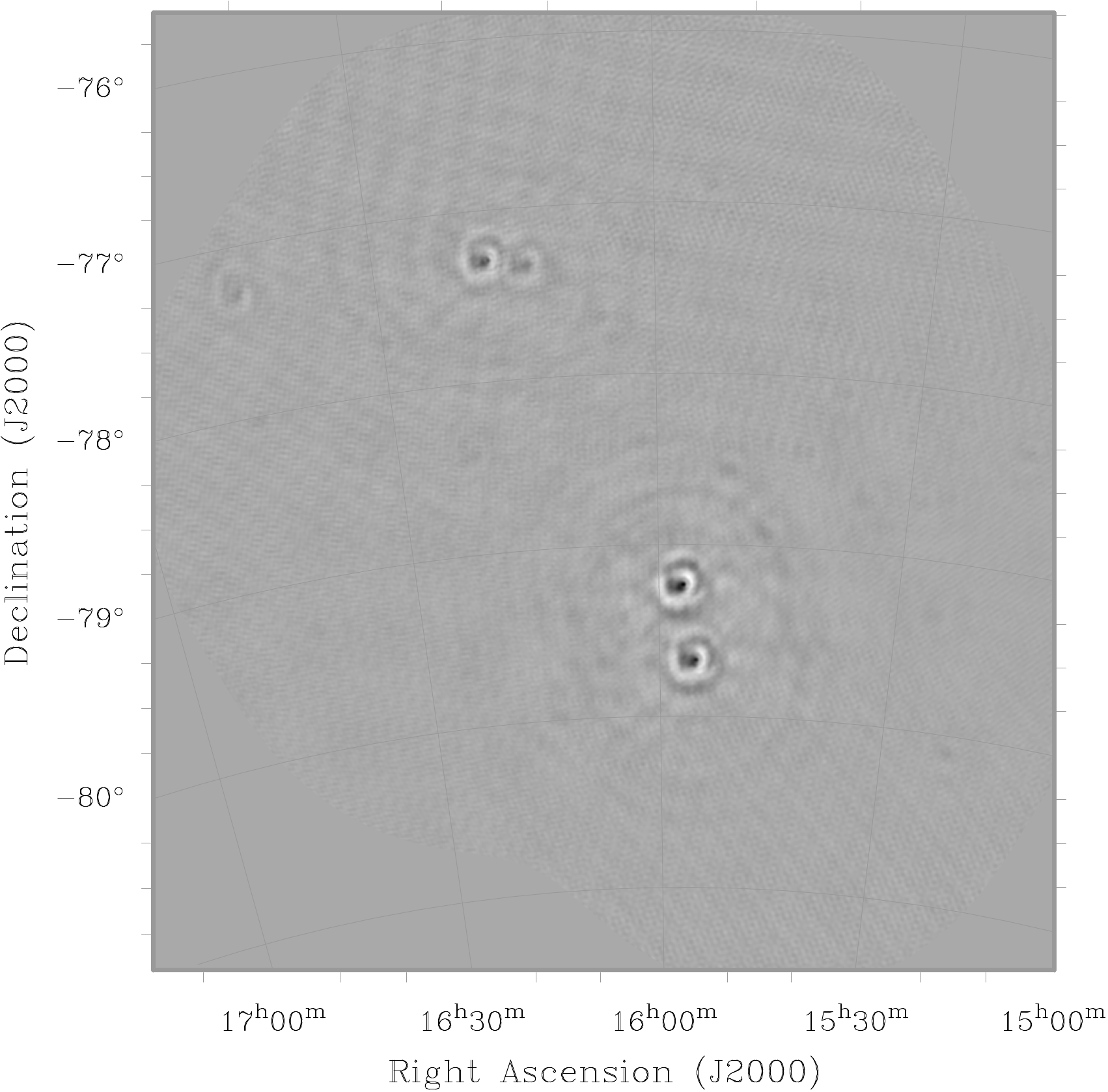}
\caption{Stokes I image formed using nine simultaneous overlapping beams and 6 baselines. 304 MHz of bandwidth (712-1016\,MHz) was recorded using the BETA correlator with a channel width of 1\,MHz.}
\label{fig:mbimg}
\end{center}
\end{figure}

The first synthesis image with multiple beams obtained using data from all 6 BETA antennas (in 2$\times$3 sub-array mode) is show in Figure \ref{fig:mbimg}. The field of interest was selected because it contains three sources (PKS 1547-795, PKS 1549-790 and PKS 1610-771), each with a flux density of roughly 5\,Jy, two of which fall within a single primary beam and another that is approximately three beam widths away. PKS 1934-638 was observed with each beam to calibrate the phases of the array. The targets were all detected (along with several other sources in the field) and their positions and fluxes match well with data from the SUMSS catalog \citep{1999AJ....117.1578B}.

Further observations will allow updates to key parameters (such as antenna positions), which will iteratively improve image quality over time. Early results from the full 15-baseline BETA (in preparation) show that traditional data processing methods can already yield high-quality images, with calibration errors being the main limiting factor. There are several reasons for this, including the presence of confusing sources near our chosen calibrator, and the fact that the Sun can be detected on the shortest baseline regardless of its position in the sky with respect to the target field. These issues are side effects of the wide field of view and will require refinements to existing data processing strategies to fully overcome. Experience gained from BETA observations will feed into the development of the final ASKAP processing pipeline.

\section{CONCLUSIONS}

The BETA system has several key roles; these include development of calibration strategies for multiple beams and the ASKAP imaging pipeline, investigating the optimal beam-forming methods to meet the needs of the ASKAP survey science projects and gaining experience with remote operation of a complex system at the MRO. The knowledge gained from operating BETA will be fed back into the final configuration of ASKAP. Preliminary observations have successfully demonstrated multi-beam synthesis imaging with phased array feeds, a significant step on the path to a new generation of wide-field radio survey instruments.

\begin{acknowledgements}
The Australian SKA Pathfinder is part of the Australia Telescope National Facility which is funded by the Commonwealth of Australia for operation as a National Facility managed by CSIRO. This scientific work uses data obtained from the Murchison Radio-astronomy Observatory (MRO), which is jointly funded by the Commonwealth Government of Australia and State Government of Western Australia. The MRO is managed by the CSIRO, who also provide operational support to ASKAP. We acknowledge the Wajarri Yamatji people as the traditional owners of the Observatory site.

Parts of this research were conducted by the Australian Research Council Centre of Excellence for All-sky Astrophysics (CAASTRO), through project number CE110001020.

The authors would also like to acknowledge the large group of people who have contributed to the planning, design, construction and support of BETA and ASKAP. This includes:

Brett Armstrong, Jay Banyer, Samantha Barry, Brayden Briggs, Ettore Carretti, Frank Ceccato, Raji Chekkala, Kate Chow, Geoff Cook, Paul Cooper, Jack Dixon, Peter Duffy, Troy Elton, Alex Harding, George Hobbs, Ron Koenig, Arkadi Kosmynin, Tom Lees, Amy Loke, Stacy Mader, Neil Marston, Vincent McIntyre, Ian McRobert, Ray Moncay, Neale Morison, John Morris, Tony Mulry, Alan Ng, Wilfredo Pena, Nathan Pope, Brett Preisig, Lou Puls, Michael Reay, Ken Reeves, Victor Rodrigues, Tim Ruckley, Craig Russell, Aaron Sanders, Ken Smart and Mark Wieringa.

We would finally like to acknowledge the contributions of the ASKAP survey science teams, represented by the following group leaders:

Shami Chatterjee, John Dickey, Bryan Gaensler, Peter Hall, Tom Landecker, Martin Meyer, Tara Murphy, Elaine Sadler, Ingrid Stairs, Lister Staveley-Smith, Russ Taylor and Steven Tingay.

\end{acknowledgements}

\bibliographystyle{apj}
\bibliography{beta}



\end{document}